\documentclass[aps,prl,twocolumn,superscriptaddress,nofootinbib,longbibliography]{revtex4-2}
\usepackage[T1]{fontenc}
\usepackage[latin9]{inputenc}
\usepackage{array}
\usepackage{multirow}
\usepackage{graphicx}
\usepackage{amsmath,amsfonts,amssymb,color}
\usepackage{amsthm}
\usepackage{mathtools}
\usepackage{leftidx}
\usepackage{xcolor}
\usepackage{dcolumn}
\usepackage{bm}
\usepackage{epstopdf}
\usepackage{epsfig}
\usepackage{environ}
\usepackage{pdfcomment}
\usepackage{float}
\usepackage[T1]{fontenc}
\usepackage[latin9]{inputenc}
\usepackage{setspace}
\usepackage{esint}
\hypersetup{colorlinks=true,citecolor=blue,
linkcolor=blue,urlcolor=blue,pdfstartview=FitH,
bookmarksopen=true}

\begin{document}

\title{Kardar-Parisi-Zhang Physics in the Density Fluctuations \\of Localized Two-Dimensional Wave Packets
}
\author{Sen Mu}\email{senmu@u.nus.edu}
\affiliation{Department of Physics, National University of Singapore, Singapore 117542, Singapore}
\affiliation{Centre for Quantum Technologies, National University of Singapore, Singapore 117543, Singapore}
\affiliation{MajuLab, CNRS-UCA-SU-NUS-NTU International Joint Research Unit, Singapore}

\author{Jiangbin Gong}\email{phygj@nus.edu.sg}
\affiliation{Department of Physics, National University of Singapore, Singapore 117542, Singapore}
\affiliation{Centre for Quantum Technologies, National University of Singapore, Singapore 117543, Singapore}
\affiliation{MajuLab, CNRS-UCA-SU-NUS-NTU International Joint Research Unit, Singapore}
\affiliation{Joint School of National University of Singapore and Tianjin
University, International Campus of Tianjin University, Binhai New City,
Fuzhou 350207, China}

\author{Gabriel Lemari\'e}\email{lemarie@irsamc.ups-tlse.fr}
\affiliation{Department of Physics, National University of Singapore, Singapore 117542, Singapore}
\affiliation{Centre for Quantum Technologies, National University of Singapore, Singapore 117543, Singapore}
\affiliation{MajuLab, CNRS-UCA-SU-NUS-NTU International Joint Research Unit, Singapore}
\affiliation{Laboratoire de Physique Th\'{e}orique, Universit\'{e} de Toulouse, CNRS, UPS, France}

\begin{abstract}
{
We identify the key features of Kardar-Parisi-Zhang universality class in the fluctuations of the wave density logarithm,  in a two-dimensional Anderson localized wave packet. In our numerical analysis, the fluctuations are found to exhibit an algebraic scaling with distance characterized by an exponent of $1/3$, and a Tracy-Widom probability distribution of the fluctuations. Additionally, within a directed polymer picture of KPZ physics, we identify the dominant contribution of a directed path to the wave packet density and find that its transverse fluctuations are characterized by a roughness exponent $2/3$. Leveraging on this connection with KPZ physics, we verify that an Anderson localized wave packet in 2D exhibits a stretched-exponential correction to its well-known exponential localization.}
\end{abstract}

\maketitle

{\it Introduction.--} Universality of fluctuations, a remarkable phenomenon pervading physics, is exemplified by the central limit theorem, which characterizes the convergence to a normal distribution for the sum of independent random variables, and describes {for instance} the behavior of particles in Brownian motion~\cite{fischer2011ctlhistory,Ornstein1930bw,Uhlenbeck1945bw}. Another example is the Kardar-Parisi-Zhang (KPZ) physics, a universal framework {which is relevant to} diverse processes ranging from interface growth to directed polymers~\cite{kpz1986, Corwin2012kpz,Takeuchi2018kpz,Spohn2020kpz}. While initially associated with classical systems, recent numerical and experimental investigations have revealed KPZ physics in quantum systems, including one-dimensional quantum magnets~\cite{Prosen2019prl,Scheie2021np,Bloch2022kpz}, random unitary circuits~\cite{Nahum2017prx}{, and driven-dissipative quantum fluids~\cite{Fontaine2022nature}.}

Anderson localization is a phenomenon where the wave function of a particle becomes localized due to disorder, hindering its diffusion~\cite{Anderson_prl1958,Abrahams_1979,Abrahams_50al}. Universal fluctuations play a key role in different aspects of this phenomenon, including universal conductance fluctuations~\cite{Lee_1985ucf,Mello_1988prl}, random matrix statistics~\cite{Beenakker_1997rmt}, log-normal distributions in one-dimensional localized systems~\cite{Mirlin_pr2000} and multifractal statistics in the critical regime of the Anderson transition~\cite{Lee_rmp1985,Evers_rmp2008, Wegner1980,Castellani_1986,Feigelman2010} {of diverse observables.}

{While one-dimensional Anderson localization is exactly solvable~\cite{Beenakker_1997rmt,Evers_rmp2008,Gogolin1976electron,Gogolin1975conductivity,Efetov1983kinetics,Hainaut2022}, understanding higher-dimensional cases and fluctuations in the localized regime remains a challenge. The forward scattering approximation and the analogy with directed polymers suggest a connection to KPZ physics in two dimensions~\cite{Stern1973,NSS1985tunnel,Kardar1992,Antonello2016,Kardar_prl1987,Zhang1995dp,comets2017dp}. Numerical simulations have confirmed this analogy for one observable, namely the conductance at zero temperature, exhibiting fluctuations belonging to the KPZ universality class~\cite{Prior_2005prb,Somoza_prl2007,Prior2009,PhysRevB.91.155413} and displaying glassy properties akin to directed polymers~\cite{Gabriel_prl2019}. However, experimental observations of electron transport in the strongly localized regime are influenced by temperature and interactions~\cite{dobrosavljevic2012conductor, pollak2013electron}, and the predicted KPZ physics in the non-interacting, zero-temperature limit has not been observed in these experiments, despite the presence of glassy phenomena such as ageing~\cite{davies1982electron,pollak1991hopping,Pollak_prl2000,Pollak_prl2008, ladieu1993conductance, PhysRevB.53.973, PhysRevLett.88.236401, PhysRevLett.99.046405}.}

Recent experimental studies on Anderson localization utilizing cold atoms, light waves and ultrasounds~\cite{Raizen_prl1994,Raizen_prl1995,Roati2008,Billy2008,Aspect2009,Palencia2010,Schwartz2007,Segev2013,Maynard2001,Hu2008,Delande_prl2015} have shed light on another type of transport, namely the expansion of a wave packet. This fresh perspective offers an in-situ and dynamical depiction of localization that differs from the conductance. Furthermore, localization in these platforms can be finely controlled, allowing for examination of a regime in which interaction and temperature effects are negligible. For example, cold atoms have provided an experimental confirmation of a three-dimensional Anderson metal-insulator transition, with a critical exponent that aligns with numerical predictions~\cite{Garreau_prl2008,Delande_prl2012}. We are thus motivated to study the universality of directly measurable  fluctuations in the spatial profile of wave packets in Anderson localized systems, in connection with the KPZ physics.

\begin{figure}
\includegraphics[width=1.0\linewidth]{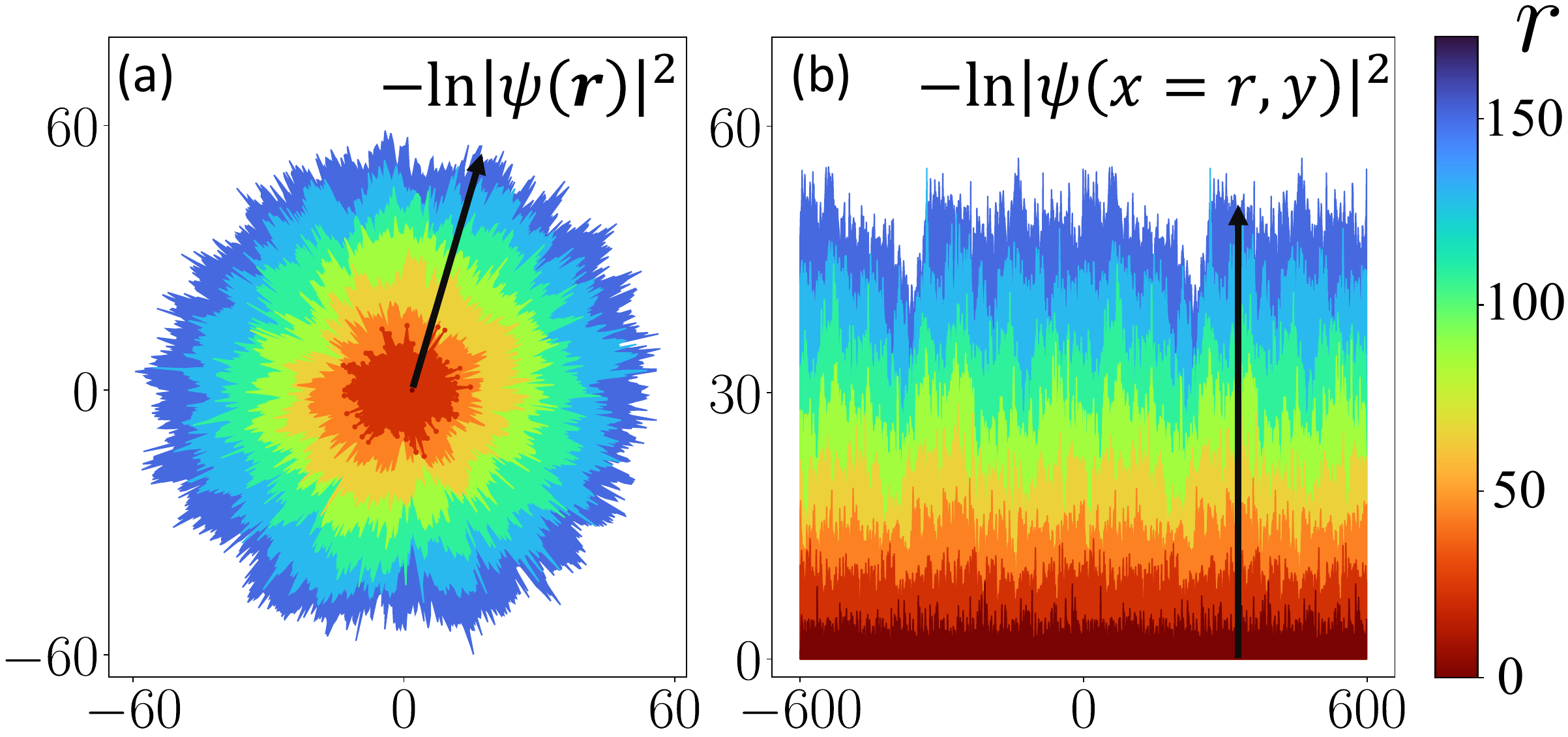}
\caption{
{The logarithm of the wave density $\ln|\psi(\boldsymbol{r})|^2$ of a 2D localized wave packet exhibits similar growth with distance $r$ as a rough surface in the Kardar-Parisi-Zhang universality class~\cite{Takeuchi2011}. We present the results of long-time evolution for two different initial conditions: (a) a ``circular'' peak at $\boldsymbol{0}$, Eq.\eqref{Eq:inicirc}, and (b) a ``flat'' line along $y$ at $x=0$, Eq.\eqref{Eq:iniflat}. We plot the transformed coordinates $(-\ln|\psi(\boldsymbol{r})|^2\frac{x}{r},-\ln|\psi(\boldsymbol{r})|^2\frac{y}{r})$ in (a) and $(y,-\ln|\psi(x=r,y)|^2)$ in (b), with different colors representing various distances $r$. Numerical simulations were conducted using model Eq.~\eqref{Eq:quantum_map}, a variant of the kicked rotor, with system size $600\times 600$, kick strength $K=1.21$, coupling $\epsilon=0.1$, and evolution time $t=10^4$.}
}
\label{fig:surface_height}
\end{figure}

We examine the fluctuations of the wave density of an exponentially localized wave packet in two dimensions. By establishing a mapping to a directed polymer model, we reveal that the fluctuations of the logarithm of the wave density correspond to the height of a rough surface in the KPZ universality class~\cite{Takeuchi2011}, see Fig.~\ref{fig:surface_height}. We find that these fluctuations scale algebraically with distance, with a fluctuation exponent of $1/3$, and identify the dominant contributing paths and their transverse fluctuations characterized by a roughness exponent of $2/3$. Moreover, we demonstrate that the distribution of the logarithm of the wave density follows the Tracy-Widom distribution. These findings firmly establish that two-dimensional localized wave packets belong to the KPZ universality class. Leveraging the well-established analytical knowledge of KPZ physics~\cite{PhysRevLett.104.230602,Calabrese_2010,Dotsenko_2010, PhysRevLett.106.250603,PhysRevE.101.012134,Spohn2020kpz} offers valuable new perspectives on the intricate characteristics of Anderson localization in two dimensions. It not only sheds light on the underlying mechanisms but also reveals intriguing features, such as the presence of a stretch-exponential correction to the exponential behavior of localization in two dimensions.

{\it Model.--} 
{In order to describe the unitary dynamics of a wave packet in a two-dimensional discrete square lattice, we} have used a variant of the quantum kicked rotor~\cite{Casati1979,Fishman1982,Grempel1982,Fishman1984,Fishman1988qkr2d} described by the following quantum map:
\begin{eqnarray}
 \vert \psi_{t+1} \rangle &=& \hat U \vert \psi_{t} \rangle = e^{-iKV(\hat{\boldsymbol k})}e^{-iW(\hat{\boldsymbol r})}  \vert \psi_{t} \rangle \; .
 \label{Eq:quantum_map}
\end{eqnarray}
This quantum map evolves the wave packet state $\vert \psi_t \rangle$ at time $t$ to the state at time $t+1$ by applying an evolution operator $\hat U$ written as the product of an operator $e^{-iW(\hat{\boldsymbol r})}$ of random on-site phases and a kick operator $e^{-iKV(\hat{\boldsymbol k})}$ playing the role of hopping amplitudes in the Anderson model~\cite{Fishman1982}.
Here $\boldsymbol r = (x,y)$ denotes site position on the discrete lattice and $\boldsymbol k = (k_x,k_y)$ with $k_x,k_y\in[-\pi,\pi)$, the wave vector reciprocal to the lattice: $\psi_t(\boldsymbol k)=\sum_{\boldsymbol r}\psi_t(\boldsymbol r)e^{-i\boldsymbol k \cdot \boldsymbol r}$. The random on-site phases $W(\boldsymbol r)$ are independent, identically distributed uniformly in the interval $[-\pi,\pi]$, and the kick operator $e^{-iKV(\hat{\boldsymbol k})}$ is parameterized by the kicking strength $K$ and $V(\boldsymbol k)=\cos k_x+\cos k_y+\epsilon\cos k_x\cos k_y$ with $\epsilon$ a nonzero coupling parameter. The quantum kicked rotor has been instrumental in understanding phenomena such as Anderson localization and the Anderson transition, both theoretically and experimentally~\cite{Casati1979,Fishman1982,Grempel1982,Fishman1984,Fishman1988qkr2d,Raizen_prl1994,Raizen_prl1995,Garreau_prl2008,PhysRevLett.105.090601,Delande_prl2015,Sanku2022}.

In this work, the initial condition of the wave packet plays an important role. {We consider two types: peaked either at a particular site on the square lattice, Eq.~\eqref{Eq:inicirc}, or a line on the lattice, Eq.~\eqref{Eq:iniflat}. The two initial conditions are  referred to as circular and flat initial conditions, respectively, in the context of KPZ physics \cite{Takeuchi2011}.} More specifically, such two types of initial states are: 
\begin{eqnarray}
{\rm circular}: \; \psi_0(\boldsymbol r) &=& \delta_{\boldsymbol r,\boldsymbol 0},\ \ r\equiv\sqrt{x^2+y^2}, \label{Eq:inicirc}\\
{\rm flat}: \;\psi_0(x,k_y) &=& \delta_{x,0} \; \delta(k_y),\ \  r\equiv x\;, \label{Eq:iniflat}
\end{eqnarray}
where $r$ represents the distance from the initial condition.

{\it Analogy with the directed polymer problem.--} 
{The directed polymer (DP) problem in $(1+1)$ dimensions is a rare example as an analytically solvable model belonging to the KPZ universality class, see e.g.~\cite{Calabrese_2010,Dotsenko_2010,PhysRevE.101.012134}. {Here, we map} our system to a variant of DP in the limit of strong disorder, which corresponds to a weak kicking strength ($K \ll 1$). Starting with a circular initial condition, Eq.~\eqref{Eq:inicirc}, we express the wave packet $\psi_t(\boldsymbol r)$ using a path integral representation:
\begin{eqnarray}
\psi_t(\boldsymbol r) =\langle \boldsymbol r| \hat U^t| \boldsymbol 0 \rangle &=& \sum_{\boldsymbol r_{t-1}}...\sum_{\boldsymbol r_{1}}\langle \boldsymbol r|\hat U|\boldsymbol r_{t-1}\rangle...\langle \boldsymbol r_1|\hat U|\boldsymbol 0\rangle.
\label{Eq:pathint}
\end{eqnarray}
When $K \ll 1$, hoppings are mainly limited to nearest neighbors with a small amplitude {$\vert J_0 \vert \ll 1$}. Consequently, we can approximate the path integral for $\psi_t(\boldsymbol r)$ by keeping only the shortest, directed paths denoted $D\mathcal P$. {In the limit of large times, we find (see SM~\cite{SuppMat}):
\begin{equation}
\psi_t(\boldsymbol r) 
 {\approx J_0^{r}} \sum_{D\mathcal{P}}\prod _{\boldsymbol r_j \in D\mathcal{P}}e^{- \tilde{W}(\boldsymbol r_j)}.
\label{Eq:fas_model}
\end{equation}
In this equation, $\tilde{W}(\boldsymbol r_j)$ are complex numbers which have a direct relationship with $W(\boldsymbol r_j)$.
The resulting expression resembles the partition function of a DP with a complex on-site disorder. Numerical evidence shows that the scaling properties of such complex DP problem are identical to those with real disorder,} see~\cite{Zhang_prl_complex_dp,Zhang_epl_complex_dp,Medina1989,GELFAND199167,BLUM1992588,Roux_1994,PhysRevLett.111.026801}, confirming their belonging to the KPZ universality class. In the next section, we will extend our analysis beyond the strong disorder regime to further confirm this connection.}

\begin{figure}
\includegraphics[width=0.8\linewidth]{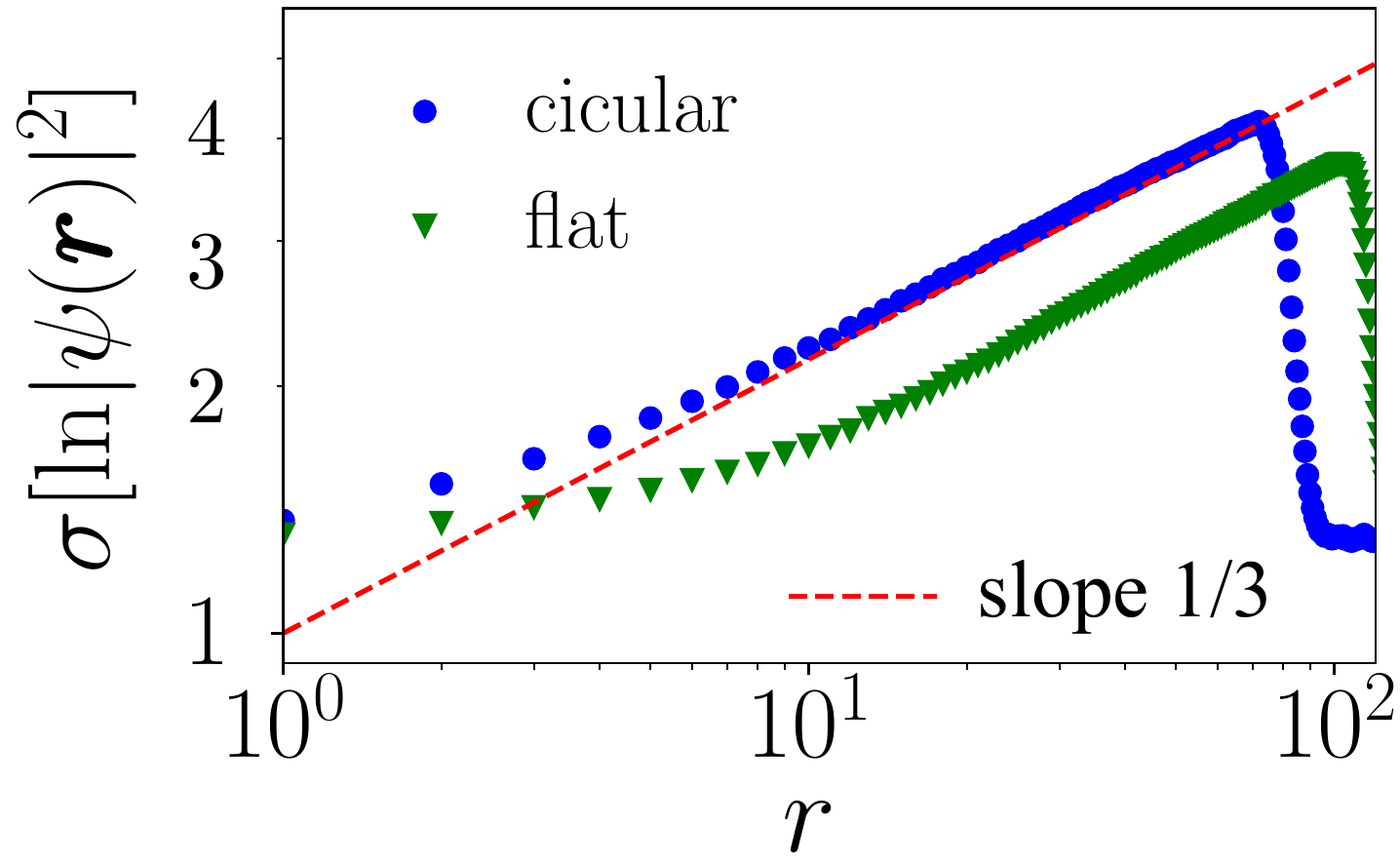}
\caption{
{
We examine the scaling of fluctuations in $\ln|\psi(r)|^2$ with distance $r$ {from} two initial conditions: (a) circular initial condition, Eq.~\eqref{Eq:inicirc}, and (b) flat initial condition, Eq.~\eqref{Eq:iniflat}. The standard deviation $\sigma[\ln|\psi(\boldsymbol r)|^2]$ is plotted along the diagonal of the square lattice $x=y$ (with distance $r=\sqrt{x^2+y^2}$) in (a), and along the line along $x$ at $y=0$ (with distance $r=x$) in (b). The dashed line represents the expected algebraic behavior $\sigma[\ln|\psi(\boldsymbol r)|^2]\sim r^{1/3}$ based on the analogy with KPZ physics. Numerical simulations are performed on system Eq.~\eqref{Eq:quantum_map} with size $300\times 300$, kick strength $K=1.04$, coupling $\epsilon=0.1$, and evolution time $t=10^3$. For both initial conditions, $10^5$ disorder realizations are considered.}
}
\label{fig:std_scaling}
\end{figure}

{\it Exponential localization as a rough surface growth.--} Let us investigate the spatial properties of the wave density $|\psi_t(\boldsymbol r)|^2$ after evolving our initial state, Eq.~\eqref{Eq:inicirc} or Eq.~\eqref{Eq:iniflat}, {for times $t \gg t_{\rm loc}$ much larger than the localization time (see SM~\cite{SuppMat}), when the envelope of the wave packet is stationary.}
We will postpone the discussion on the effects of time and omit the label of time for the wave density in the following, i.e. $|\psi(\boldsymbol{r})|^2 \equiv |\psi_t(\boldsymbol{r})|^2$. Consider the logarithm of the wave density $\ln|\psi(\boldsymbol r)|^2$ at a distance $r$ from the initial location. The analogy that we have drawn in the previous section leads us to interpret $-\ln|\psi(\boldsymbol r)|^2$ as {the analogue of} the height function $h$ of a growing rough surface (or, equivalently, as the free energy of a directed polymer problem), where the distance $r$ in our model is understood as the time in the surface growth (or directed polymer) process. In Fig.~\ref{fig:surface_height}, we present this effective growth for both circular, Eq.~\eqref{Eq:inicirc}, and flat, Eq.~\eqref{Eq:iniflat}, initial conditions. The striking resemblance between these plots and experimental observations in liquid crystal nematics~\cite{Takeuchi2011} will be quantitatively validated in the following.

Based on the analogy with KPZ physics~\cite{PhysRevB.43.10728,Takeuchi2011, monthus2012random,Somoza_prl2007} and extending results from Anderson localization in one dimension~\cite{Mirlin_pr2000}, we expect the logarithm of the wave packet density $|\psi(\boldsymbol r)|^2$ to behave as follows:
\begin{equation}
\ln|\psi(\boldsymbol r)|^2 \approx -\frac{2r}{\xi}+\left(\frac{r}{\xi}\right)^\beta \Gamma \, \chi(\boldsymbol{r}),
\label{Eq:stat_dis}
\end{equation}
where $\Gamma$ is a constant, and $\chi$ is a random variable of order one. The first term {corresponds to exponential localization}, with $\xi$ the localization length, while the second term captures fluctuations with a fluctuation growth exponent $\beta$. In two dimensions, $\beta$ differs from the known value ($\beta = 1/2$) for Anderson localization in one dimension~\cite{Mirlin_pr2000}. Fig.~\ref{fig:std_scaling} shows the behavior of the standard deviation $\sigma[\ln|\psi(\boldsymbol r)|^2]$ of $\ln|\psi(\boldsymbol r)|^2$, which grows algebraically with $r$ as $\sigma \sim r^{\beta}$. The fluctuation exponent $\beta \approx 1/3$, consistent with the KPZ universality class.

\begin{figure}
\includegraphics[width=1\linewidth]{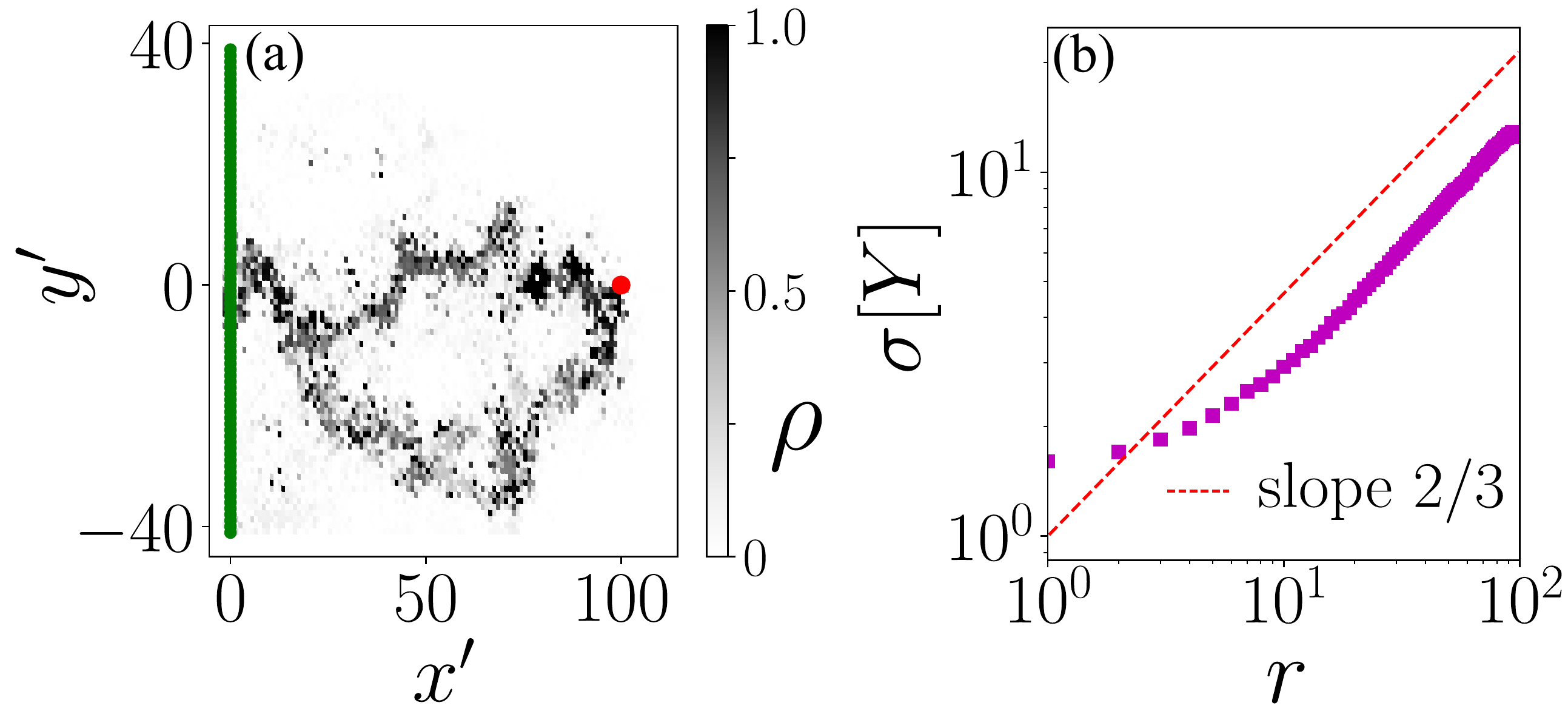}
\caption{(a) The optimal path for a wave packet originating from a flat initial condition, Eq.~\eqref{Eq:iniflat}, represented by the green line along $y$ at $x=0$, propagates to the final site $\boldsymbol r = (x=100,y=0)$ shown as the red dot. %{The transverse position $Y$, Eq.~\eqref{eq:avy}, is at the magenta diamond}. 
The local response $\rho_{\boldsymbol r}(\boldsymbol r^\prime)$, Eq.~\eqref{eq:rhorprime}, is displayed as a colorplot, with $\rho_{\boldsymbol r}(\boldsymbol{r}^\prime)=1$ if $\rho_{\boldsymbol r}(\boldsymbol{r}^\prime)>1$ for better visibility.
(b) The scaling of the standard deviation $\sigma[Y]$ of the transverse position $Y$, Eq.~\eqref{eq:avy}, of the optimal path is shown as a function of the distance $r=x$ from the flat initial condition, Eq.~\eqref{Eq:iniflat}. The dashed line represents the expected algebraic behavior $\sigma[Y] \sim r^{2/3}$ based on KPZ physics.
The numerical simulations employ the model described by Eq.~\eqref{Eq:quantum_map} with a system size of $300\times 300$, kick strength $K=1.04$, coupling $\epsilon=0.1$, evolution time $t=10^3$, and an averaging over $10^4$ disorder realizations.}
\label{fig:dpath_roughness}
\end{figure}

{\it Optimal path and roughness exponent.--} 
{The optimal trajectory of a directed polymer in a random medium reveals important insights into the system's dynamics and glassy properties~\cite{mezard1990glassy,PhysRevB.43.10728,derrida1988polymers}. By extending the approach in~\cite{Gabriel_prl2019}, we can visualize the optimal path associated with a wave packet localized in a specific disorder configuration. This is achieved by examining the response of the wave density at $\boldsymbol r$ to a local perturbation of the disorder at $\boldsymbol r^\prime$, where the perturbation involves shifting the on-site phase $W(\boldsymbol r^\prime)$ by $\pi$. We quantify this response using 
\begin{equation}
\rho_{\boldsymbol r}(\boldsymbol r^\prime) \equiv \frac{\left\vert{|{\tilde \psi}_{\boldsymbol r^\prime}(\boldsymbol{r})|^2-|{\psi}(\boldsymbol{r})|^2}\right\vert}{|{\psi}(\boldsymbol{r})|^2},
\label{eq:rhorprime}
\end{equation}
which measures the difference between the wave densities with or without perturbation {($|{\tilde \psi}_{\boldsymbol r^\prime}(\boldsymbol{r})|^2$ or $|{\psi}(\boldsymbol{r})|^2$, respectively)}. In the strongly localized regime ($r\gg \xi$), the response becomes highly inhomogeneous, concentrated along a specific path that depends on both the disorder configuration, initial condition, and final point $\boldsymbol r$. In Fig.~\ref{fig:dpath_roughness}(a), we illustrate such a path, where the red dot represents the final site $\boldsymbol r$, and the green line corresponds to the flat initial condition along the line $x=0$ (Eq.~\eqref{Eq:iniflat}).}

{Let us delve deeper into the properties of the optimal path. Considering the same flat initial condition, we focus on the transverse position $Y$ at $x=0$ of the optimal path for the wave packet at $\boldsymbol r = (x,0)$. The calculation of $Y$ relies on the response $\rho_{\boldsymbol r}$ at $x'=0$, given by
\begin{equation}
Y = \dfrac{\sum_{y'}y' \,\rho_{\boldsymbol r}(0,y')}{\sum_{y'}\rho_{\boldsymbol r}(0,y')},
\label{eq:avy}
\end{equation}
where $\rho_{\boldsymbol r}(0,y')/\sum_{y'}\rho_{\boldsymbol r}(0,y')$ represents the normalized probability distribution of the optimal trajectory. In the directed polymer problem, the optimal path arises from a global optimization over the disorder, resulting in a path that exhibits more wandering compared to a simple random walk. This wandering behavior is characterized by the roughness exponent $\zeta$, which relates the standard deviation of the expected transverse position to the distance $r$ from the initial condition (Eq.~\eqref{Eq:iniflat}): $\sigma[Y] \sim r^{\zeta}$. Numerical results on the roughness exponent are presented in Fig.~\ref{fig:dpath_roughness}(b), clearly indicating $\zeta \approx 2/3$. This finding reinforces the conclusion that wave packet localization in two dimensions belongs to the KPZ universality class.}

\begin{figure}
\includegraphics[width=1.0\linewidth]{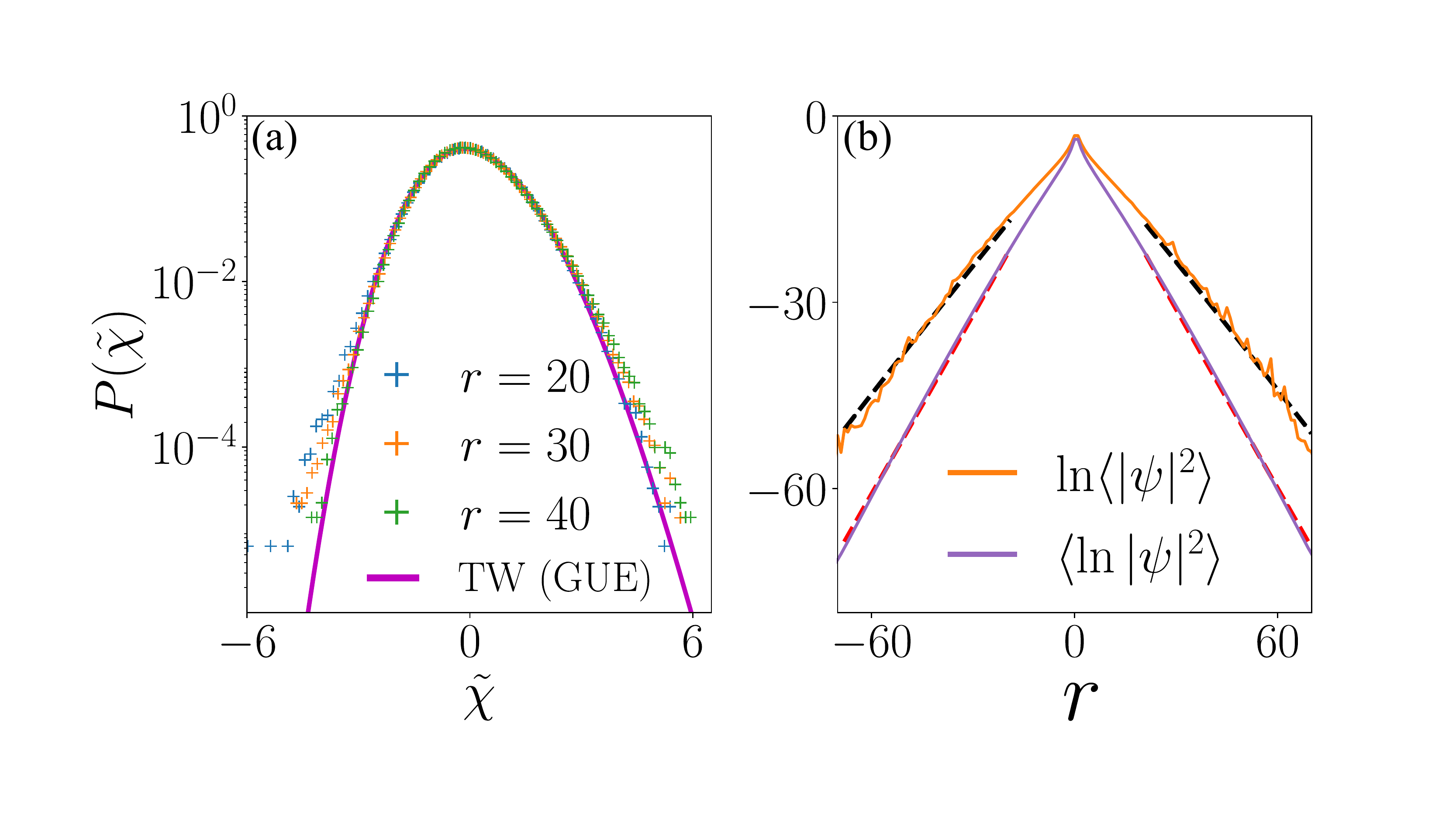}
\caption{{Tracy-Widom distribution and the shape of a 2D localized wave packet.
(a) The distribution of the rescaled wave density $\tilde{\chi}$, defined as $(\ln|\psi(\boldsymbol r)|^2-\langle\ln|\psi(\boldsymbol r)|^2\rangle)/{\sigma[\ln|\psi(\boldsymbol r)|^2]}$, exhibits good agreement with the GUE Tracy-Widom distribution (magenta line) for the circular initial condition, Eq.~\eqref{Eq:iniflat}. Crosses of different colors represent different distances $r$ from the initial condition, and their collapse onto the Tracy-Widom distribution confirms the scaling behavior given by Eq.~\eqref{Eq:stat_dis}.
(b) The typical wave density $\langle\ln|\psi(r)|^2\rangle$ (lower violet line) and the logarithm of the average wave density $\ln\langle|\psi(r)|^2\rangle$ (upper orange line) are displayed. The red dashed line represents a fit using Eq.~\eqref{Eq:fit_ave_log} with two fitting parameters: $\xi\approx 2.1$ and $\Gamma\approx 0.7$. The upper black dashed line represents Eq.~\eqref{Eq:ave_den} with $\Gamma^\prime\approx -6.8$, $\Gamma^{\prime\prime}\approx 2.7$, and $\Gamma^{\prime\prime\prime}\approx 5.3$, determined from the typical wave density and the GUE Tracy-Widom distribution.
Numerical data is obtained from the model described by Eq.~\eqref{Eq:quantum_map}, considering $\boldsymbol r$ on the diagonal $x=y$ of a square lattice with size $300\times 300$. The simulations employ a kick strength of $K=1.0$, coupling $\epsilon=0.1$, an evolution time of $t=10^3$, and involve $10^6$ disorder realizations.}
}
\label{fig:tw_wave_density}
\end{figure}

{\it Tracy-Widom distribution and the shape of a 2D localized wave packet.--} 
Drawing from the analogy with the DP problem, the random variable $\chi$ in Eq.~\eqref{Eq:stat_dis} is expected to follow a universal distribution function. Specifically, for the circular initial condition, it should obey the Tracy-Widom (TW) distribution of the Gaussian unitary ensemble (GUE). In Fig.~\ref{fig:tw_wave_density}(a), we present the distribution of $\tilde{\chi}(\boldsymbol r)=(\ln|\psi(\boldsymbol r)|^2-\langle\ln|\psi(\boldsymbol r)|^2\rangle)/{\sigma[\ln|\psi(\boldsymbol r)|^2]}$ alongside the GUE TW distribution $P_{\rm TW}(\tilde \chi)$, rescaled to have zero mean and unit standard deviation. We observe a good agreement {between the two}.

{This} makes it possible to determine the shape of a localized wave packet. The form of the typical wave density follows from an average of the result in Eq.~\eqref{Eq:stat_dis}:
\begin{equation}
\langle\ln|\psi(\boldsymbol r)|^2\rangle \displaystyle \underset{r\gg \xi}{\approx} -\frac{2r}{\xi}+\left(\frac{r}{\xi}\right)^{1/3} \Gamma \, \mu\; ,
\label{Eq:fit_ave_log}
\end{equation}
where $\langle...\rangle$ denotes disorder averaging at fixed $r$ and $\mu\approx -1.77$ is the non-zero mean of the GUE TW distribution. 
Fig.~\ref{fig:tw_wave_density}(b) shows a very good agreement between this prediction Eq.~\eqref{Eq:fit_ave_log} and numerical data for $\langle \ln|\psi(\boldsymbol r)|^2\rangle$ at large {$r \gg \xi$}. 

{The average wave density $\langle|\psi(r)|^2\rangle$ can be determined by considering the typical wave packet and the fluctuations around it, following the GUE TW distribution $P_{\rm TW}(\tilde{\chi})$. This average wave packet profile is experimentally accessible, for instance, with cold atoms, see e.g.~\cite{PhysRevLett.105.090601,Delande_prl2015,Billy2008}. After some algebraic manipulations (see SM~\cite{SuppMat} for details), we obtain the approximation:
\begin{eqnarray}
\langle|\psi(r)|^2\rangle &\underset{r\gg \xi}{\approx}& e^{-\frac{2r}{\xi}+(\frac{r}{\xi})^{1/3} \Gamma \,\mu}\int_{-\infty}^{\infty}{\rm d}\tilde{\chi} e^{\tilde{\chi} \sigma[\ln|\psi(\boldsymbol r)|^2]}P_{\rm TW}(\tilde{\chi})\nonumber\\
&\approx& e^{-\frac{2r}{\xi}+(\frac{r}{\xi})^{1/3} \Gamma^\prime+(\frac{r}{\xi})^{2/3} \Gamma^{\prime\prime}+\Gamma^{\prime\prime\prime}},
\label{Eq:ave_den}
\end{eqnarray}
where $\sigma[\ln|\psi(\boldsymbol r)|^2] \sim r^{1/3}$, and new constants $\Gamma^\prime$, $\Gamma^{\prime\prime}$, and $\Gamma^{\prime\prime\prime}$ are introduced. Fig.~\ref{fig:tw_wave_density}(b) compares this prediction with numerical data for $\langle|\psi(r)|^2\rangle$. Notably, the values of $\Gamma^\prime$, $\Gamma^{\prime\prime}$, and $\Gamma^{\prime\prime\prime}$ are not fitted from $\langle|\psi(r)|^2\rangle$, but instead determined from the typical wave packet and the GUE TW distribution {$P_{\rm TW}(\tilde{\chi})$.}

The excellent agreement between Eqs.~\eqref{Eq:fit_ave_log}-\eqref{Eq:ave_den} and the numerical data predicts a stretch-exponential correction to the exponential behavior of 2D localization. Notably, the absence of this stretch-exponential correction in 1D Anderson localization, where Gaussian fluctuations of the logarithm of the wave density have zero average~\cite{Mirlin_pr2000}, underscores its distinctive signature in 2D {localization}.}

{\it Conclusion.--}
In this Letter, we have unveiled previously unknown and highly significant features of two-dimensional Anderson localization of wave packets through an analogy with KPZ physics. Notably, we have discovered that the fluctuations of the logarithm of the wave density exhibit an algebraic growth with the KPZ growth exponent of 1/3, providing a stretch exponential correction to the prevailing exponential localization behavior in two dimensions. These findings have direct experimental accessibility with cold atom systems and classical waves~\cite{Raizen_prl1994,Billy2008,Garreau_prl2008,PhysRevLett.105.090601,Hainaut2022,Hu2008,Segev2013,lagendijk2009fifty,Delande_prl2015,Schwartz2007}, opening up exciting opportunities for experimental validation and further exploration. Moreover, our extensive numerical results demonstrate that the critical exponents and statistical distributions governing KPZ physics are universally present in localized wave packets in two dimensions. {Indeed, while our computational results are based on a variant of the kicked rotor, we have confirmed the validity of our observations in the two-dimensional Anderson model (see SM~\cite{SuppMat}). We have omitted time effects from our description. In the regime of large times $t \gg t_{\text{loc}}$, the spatial distributions reach a stationary state, which implies that our findings are relevant for any fixed time $t \gg t_{\text{loc}}$. However, in addition to the spatial perspective, the wave packet observable also provides insight into the temporal fluctuations. This opens up intriguing possibilities for investigating the dynamical glassy properties inherent in the DP problem \cite{wiese2022theory}, including phenomena such as aging \cite{yoshino1996off, PhysRevE.55.5651}.}
The integration of KPZ and directed polymer physics insights holds also great promise in shedding light on the as-yet-unclear aspects of Anderson localization in high dimensions~\cite{arenz2023wegner,garcia2022critical,monthus2006freezing,baroni2023corrections}.

{\it Acknowledgement.--} We thank B.~Georgeot, N.~Izem, P.~Le Doussal, C.~Miniatura, M.~Richard and G.~Schehr for fruitful discussions, and M.~Richard for a careful reading of the manuscript. This study has been supported through
the research funding Grants No. ANR-17-CE30-0024, ANR-18-CE30-0017 and ANR-19-CE30-0013, and by the Singapore Ministry of Education Academic Research Fund Tier I (WBS
No. R-144-000-437-114). J.G. also acknowledges the support from the Singapore National Research Foundation via the project NRF2021-QEP2-02-P09, as well as the support by the National Research Foundation, Singapore and A*STAR under its CQT Bridging Grant.  
The computational work for this Letter was performed on resources of the National Supercomputing Centre, Singapore and Calcul en Midi-Pyr\'en\'ees (CALMIP), France.

%%%%%%%%%%%Supplementary Materials%%%%%%%%%%%%%%%%%%
\clearpage
\onecolumngrid
\begin{center}
\textbf{\large Supplemental Materials}\end{center}
\setcounter{equation}{0}
\setcounter{figure}{0}
\renewcommand{\theequation}{S\arabic{equation}}
\renewcommand{\thefigure}{S\arabic{figure}}
\renewcommand{\cite}[1]{\citep{#1}}

\section{Mapping to the directed polymer problem}

In this section, we provide a detailed derivation of the mapping to the complex directed polymer (DP) discussed in the main text, in the limit of strong disorder, which corresponds to a weak kicking strength $K\ll 1$. For simplicity, we consider the circular initial condition given by {Eq.~(2) in the main text}, and express $\psi_t(\boldsymbol r)$, the wave packet at position $\boldsymbol r$ and time $t$, as the result of a path integral:
\begin{eqnarray}
\psi_t(\boldsymbol r) &=& \sum_{\boldsymbol r_{t-1}}...\sum_{\boldsymbol r_{1}}\langle \boldsymbol r|\hat U|\boldsymbol r_{t-1}\rangle...\langle \boldsymbol r_1|\hat U|\boldsymbol 0\rangle \; .
\label{Eq:pathint-SM}
\end{eqnarray}
This expression is obtained by inserting the resolution of identity $\sum_{\boldsymbol r_j}|\boldsymbol r_j\rangle\langle \boldsymbol r_j|=\hat 1$ between each pair of consecutive evolution operators $\hat U$. In the regime where $K \ll 1$, the hopping processes are predominantly limited to nearest neighbors, and the amplitude $J_0$ of these hoppings is small, satisfying $|J_0|\ll 1$.

Let us denote $\mathcal P :=  \boldsymbol r_0 = \boldsymbol 0 \rightarrow \boldsymbol r_1 \rightarrow \cdots \rightarrow \boldsymbol r_{n-1} \rightarrow \boldsymbol r_n= \boldsymbol r $ as one of the paths of length $n$, where $\boldsymbol r_{j+1}$ is a nearest neighbor of $\boldsymbol r_{j}$ for all $0\le j\le n-1$ (which implies that $\boldsymbol r_{j+1} \ne \boldsymbol r_{j}$). The length $n$ of the path $\mathcal{P}$ corresponds to the number of nearest neighbor hoppings performed along the path. In the path integral given by Eq.~\eqref{Eq:pathint-SM}, a path of length $n$ contributes a term proportional to $J_0^n$. Since we have $t$ scattering events (i.e., $t$ operators $\hat U$ in Eq.~\eqref{Eq:pathint-SM}), the length $n$ of $\mathcal P$ cannot exceed $t$. Furthermore, considering only nearest-neighbor hoppings and starting from $\boldsymbol 0$ and ending at $\boldsymbol r$ with a lattice distance of $d$ between them, the path length must be at least $d$. Therefore, in the nearest-neighbor approximation, we must have $t \ge d$.

Note that there is a distinction between the lattice distance $d$ and the distance $r$ defined in Eqs.~(2) and (3) from the main text. In the case of the circular initial condition, $d$ corresponds to $x + y$, while $r$ represents $\sqrt{x^2 + y^2}$. However, since we are considering a scenario in the main text where the kicking strength is not always small, meaning that the hoppings are not limited to nearest neighbors, and since $d$ and $r$ scale linearly with $x$ and $y$, we can omit this distinction. Therefore, in the remainder of this section, we will replace $d$ with $r$.

When $n < t$, a path $\mathcal{P}$ of length $n$ will encounter certain points $\boldsymbol{r}_j$ and remain on them for a consecutive number of periods of time, denoted as $n_{\boldsymbol{r}_j} > 1$. This event contributes a phase of $n_{\boldsymbol{r}_j} W(\boldsymbol{r}_j)$ to the path integral. Since we have a total of $t$ scattering events, we can express $t$ as the sum of $n_{\boldsymbol{r}_j}$ over all $\boldsymbol{r}_j$ encountered along the path, i.e., $t = \sum_{j=0}^{n-1} n_{\boldsymbol{r}_j}$.

In the limit $\vert J_0 \vert \ll 1$ we are considering, the contribution of a path decreases exponentially with its length $n$ as $\vert J_0 \vert^n$. Therefore, we can apply the forward scattering approximation, which involves considering only the shortest directed paths $D\mathcal{P}$ connecting the initial and final sites. These paths do not repeat, always travel in the direction from $\boldsymbol{0}$ to $\boldsymbol{r}$, and have a length $r$, which is the lattice distance between $\boldsymbol{0}$ and $\boldsymbol{r}$ (we have replaced $d$ by $r$, as mentioned above). Thus, we obtain the following expression:
\begin{equation}
\psi_t(\boldsymbol{r}) \approx {J_0}^{r} \sum_{D\mathcal{P}} \sum_{\{n_{\boldsymbol{r}_j}\}} \prod_{j=0}^{r-1} e^{-i n_{\boldsymbol{r}_j} W(\boldsymbol{r}_j)},
\label{Eq:fas_model-SM}
\end{equation}
where the sum over $\{n_{\boldsymbol{r}_j}\}$, denoted as $\sum_{\{n_{\boldsymbol{r}_j}\}}$, is subject to the constraint $t=\sum_{j=0}^{r-1} n_{\boldsymbol{r}_j}$. However, evaluating this sum explicitly at finite time is challenging due to this constraint.

In the limit of infinite time, we can solve the sum explicitly. To do this, we transform to a frequency representation by defining:
\begin{equation}
    \psi_\omega(\boldsymbol r) \equiv \sum_{t=0}^{\infty} e^{i \omega^+ t} \psi_t(\boldsymbol{r}) \; ,
\end{equation}
with $\omega^+= \omega + i \eta$ and $0<\eta\ll 1$ a small imaginary shift of the frequency to regularize the divergences. In this case, we can rewrite the equivalent of Eq.~\eqref{Eq:fas_model-SM} for $\psi_\omega(\boldsymbol r)$ as:
\begin{equation}
    \psi_\omega(\boldsymbol r) \approx {J_0}^{r} \sum_{D\mathcal{P}} \prod _{j=0}^{r-1} \sum_{n_j=0}^{\infty} e^{i n_j [\omega^+-W(\boldsymbol r_j)]} =  {J_0}^{r} \sum_{D\mathcal{P}} \prod _{j=0}^{r-1} e^{-\tilde{W}(\boldsymbol r_j)} \; \;,
    \label{eq:FSAPsiomega}
\end{equation}
where $\tilde{W}(\boldsymbol{r}_j)$ are complex numbers defined as:
\begin{equation}
   e^{-\tilde{W}(\boldsymbol r_j)} \equiv  \frac{1}{1-e^{i [\omega^+-W(\boldsymbol r_j)]}}\;.
\end{equation}

It is conjectured that there exist similar complex numbers, denoted as $\tilde{W}(\boldsymbol r_j)$,
which allow the amplitude of a directed path $\sum_{\{n_{\boldsymbol r_j}\}} \prod_{j=0}^{r-1} e^{-i n_{\boldsymbol r_j}  W(\boldsymbol r_j)}$ for $\psi_t(\boldsymbol r)$ in Eq.~\eqref{Eq:fas_model-SM} to be expressed as:
\begin{equation}
   \sum_{\{n_{\boldsymbol r_j}\}} \prod_{j=0}^{r-1} e^{-i n_{\boldsymbol r_j}  W(\boldsymbol r_j)} =  \prod _{j=0}^{r-1} e^{-\tilde{W}(\boldsymbol r_j)} \;.
\end{equation}
These complex numbers, if they exist, are directly related to the random on-site phases $W(\boldsymbol r_j)$, and it is expected that their time dependence is slow. We can then express
\begin{equation}
    \psi_t(\boldsymbol r) 
 \approx {J_0} ^{r} \sum_{D\mathcal{P}} \prod _{j=0}^{r-1} e^{-\tilde{W}(\boldsymbol r_j)} \;.
 \label{eq:FSAPsit}
\end{equation}

Equations (\ref{eq:FSAPsit}) and \eqref{eq:FSAPsiomega} bear a resemblance to the partition function $Z$ of a directed polymer (DP) on a lattice, but with the inclusion of complex on-site energies $\tilde{W}(\boldsymbol{r}_j)$ at unit temperature. Although the complex weights introduce intriguing interference effects, numerical investigations have revealed that the scaling and fluctuation properties of these complex DPs are indistinguishable from those of directed polymers with real on-site energies, as reported in~\cite{Zhang_prl_complex_dp,Zhang_epl_complex_dp,Medina1989,GELFAND199167,BLUM1992588,Roux_1994}. Consequently, both systems belong to the KPZ universality class.

In summary, the strong disorder limit establishes an analogy between the localization of wave packets in two dimensions and the DP problem, where the amplitude of the wave packet $\vert \psi_t(\boldsymbol{r})\vert^2$ corresponds to the partition function of a complex DP. This intriguing connection suggests that the spatial fluctuations exhibited by two-dimensional localized wave packets are governed by KPZ physics.

\section{Localization time and procedure to build Figure 1 in the main text}

In this section, we give more details on the quantum dynamics of an initially peaked wave packet by employing the quantum map Eq.~(1) discussed in the main text. Our analysis starts with examining the dynamics of the inverse participation ratio (IPR) for the wave packet. This allows us to determine the localization time and subsequently focus on the spatial properties of localized wave packets in our further investigations. By studying the density profiles of these localized wave packets, we outline the procedure used to create Fig.~1 in the main text, which depicts the growth of a rough surface. 

\begin{figure}
\includegraphics[width=0.95\linewidth]{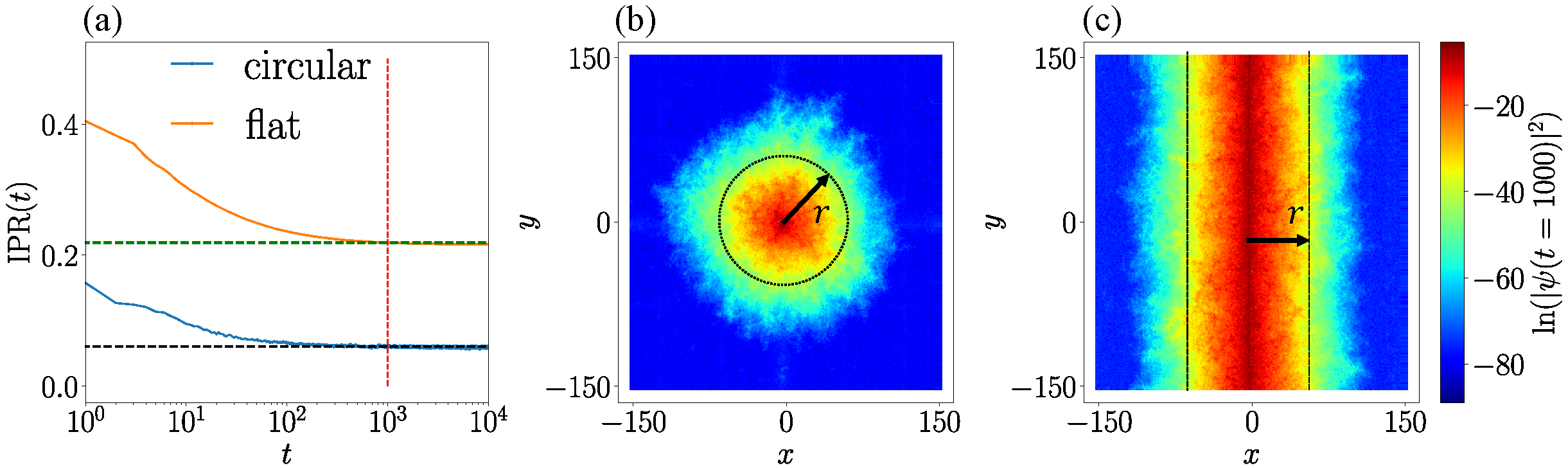}
\caption{(a) Evolution of the inverse participation ratio (IPR) for circular and flat initial conditions. The red dashed line corresponds to $t=1000$, while the green and black dashed lines represent ${\rm IPR}(t=1000)$ for the flat and circular cases, respectively. The numerical simulations were conducted on a system of size $300\times 300$ with a kick strength of $K=1.04$, coupling $\epsilon=0.1$, evolution time $t=10^4$, and 1280 disorder realizations. Density profiles of the wave packets at $t=1000$ are shown for a single disorder realization in (b) for the circular initial condition and (c) for the flat initial condition. In (b) and (c), the black dashed lines indicate the wave density $\ln|\psi(\boldsymbol r)|^2$ at a distance $r=\sqrt{x^2+y^2}$ for the circular condition and $\ln|\psi(x=r,y)|^2$ at a distance $r=x$ for the flat condition, respectively. 
%{(GL: make a .png version of that figure, which is too heavy in the .pdf version.)} {MS: done.}
}
\label{fig:ipr_density}
\end{figure}

{In a large two-dimensional disordered medium, an initially peaked wave packet undergoes exponential localization over sufficiently long time.} To estimate the localization time scale, denoted as $t_{\rm loc}$, we track the dynamics of the inverse participation ratio (IPR). We have reached the localization time once the IPR saturates, indicating the envelope of the wave packet has reached a stationary regime. The specific definition of IPR depends on the initial conditions. For the circular condition, where $\psi_0(\boldsymbol r) = \delta_{\boldsymbol r,\boldsymbol 0}$, we have ${\rm IPR}(t) = \sum_{\boldsymbol r}|\psi_t(\boldsymbol r)|^4$. For the flat condition, with $\psi_0(x,k_y) = \delta_{x,0} \; \delta(k_y)$, we calculate ${\rm IPR}(t)$ by averaging ${\rm IPR}_y = \sum_x|\psi_t(x,y)|^4/(\sum_x|\psi_t(x,y)|^2)^2$ along the line at $y$ and normalizing it by $1/N$. Fig.~\ref{fig:ipr_density}(a) shows that the IPR saturates after $t=1000$ for both initial conditions, indicated by the red dashed line, with a kick strength $K=1.0$ and coupling $\epsilon=0.1$. Thus, we choose $t=1000$ as the evolution time for our numerical simulations shown in the main text. 

Fig.~\ref{fig:ipr_density}(b) and Fig.~\ref{fig:ipr_density}(c) depict the density of localized wave packets at $t=1000$ for a single disorder realization, considering circular and flat initial conditions, respectively. To construct Fig.~1 in the main text, we collect the wave packet density at a distance $r$ from its initial launch position, indicated by black dashed lines. For the circular condition, we record the coordinates $(x, y)$ of the wave density whose distance to the origin is $r=\sqrt{x^2+y^2}$. Using the wave density $\ln|\psi(\boldsymbol r)|^2$ at $|\boldsymbol r|=r$, we create a new set of variables $(-\ln|\psi(\boldsymbol r)|^2\frac{x}{r},-\ln|\psi(\boldsymbol r)|^2\frac{y}{r})$ and plot them for different $r$. In Fig.~1(a) of the main text, the black arrow represents the magnitude of $-\ln|\psi({\boldsymbol r})|^2$ at a fixed $r$, which fluctuates with respect to ${\boldsymbol r}$. For the flat condition, we collect the wave density at $(x=r, y)$ and plot $-\ln|\psi(x=r,y)|^2$ against $y$ for different $r$. Remarkably, the quantity $-\ln|\psi(\boldsymbol r)|^2$ exhibits similar behavior to that of a height function on a growing rough surface, where the distance $r$ plays the role of time in the KPZ process.

\section{Derivation of the shape of the average wave density}

\begin{figure}
\includegraphics[width=0.95\linewidth]{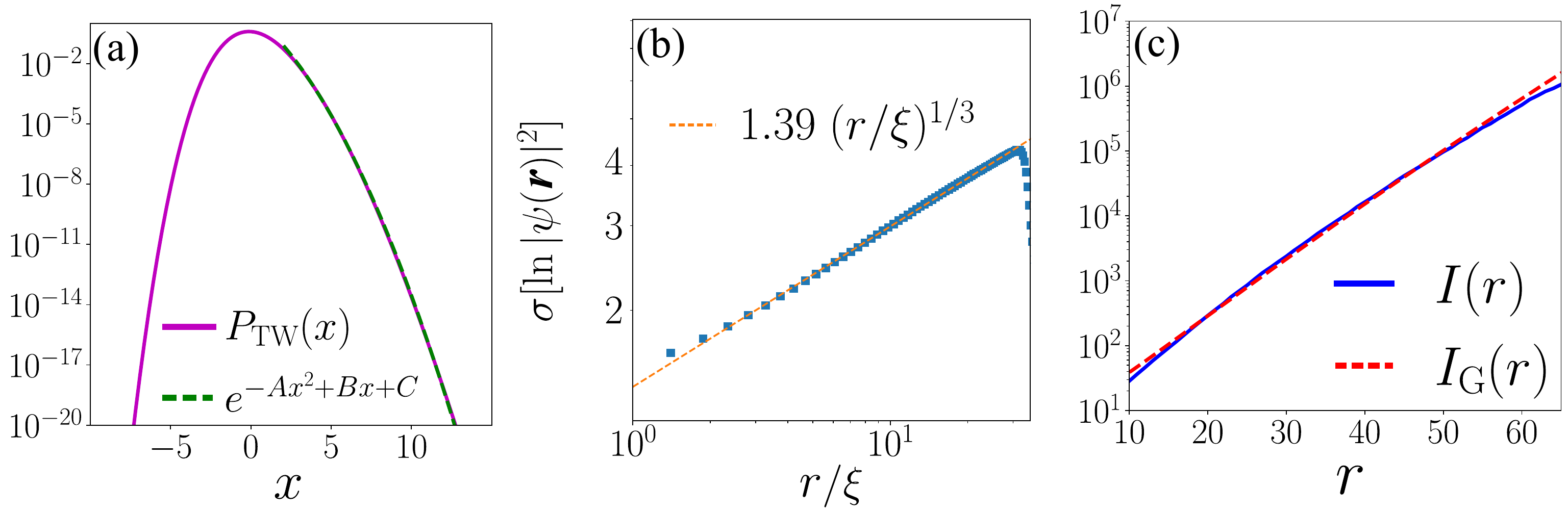}
\caption{(a) Approximation to the rescaled GUE TW distribution $P_{\rm TW}(x)$ by Eq.~\ref{Eq:tw_approx} with $A=0.18$, $B=-1.42$, and $C=1.10$ for the range $x\in [2,12]$. (b) Fitting to the scaling of fluctuations in $\ln|\psi(\boldsymbol r)|^2$ with the rescaled distance $r/\xi$ by $D(r/\xi)^{1/3}$ with $D=1.39$ for the circular initial conditions. Other parameters are the same as in Fig.~4 of the main text. (c) Comparison between the behavior of $I(r)$ and $I_{\rm G}(r)$ in Eq.~\ref{ave_den_integration} with parameters $A,B,C$ and $D$ determined above. 
}
\label{fig:average_wave_den}
\end{figure}

In this section, we provide more details on our derivation of Eq.~(10) describing the average wave density. By considering the rescaled variable $\tilde{\chi}(\boldsymbol r)=(\ln|\psi(\boldsymbol r)|^2-\langle\ln|\psi(\boldsymbol r)|^2\rangle)/\sigma[\ln|\psi(\boldsymbol r)|^2]$, which follows the rescaled GUE Tracy-Widom (TW) distribution $P_{\rm TW}$ with zero mean and unit standard deviation, we can derive an expression for the average wave density. If $P_{\rm ave}(|\psi(\boldsymbol r)|^2)$ is the distribution function for $|\psi(\boldsymbol r)|^2$, the average wave density can be obtained as follows:
\begin{equation}
\langle|\psi(\boldsymbol r)|^2\rangle = \int_0^1 |\psi(\boldsymbol r)|^2 P_{\rm ave}(|\psi(\boldsymbol r)|^2) \; {\rm d}|\psi(\boldsymbol r)|^2.
\end{equation} 
By a change of variable and utilizing the distribution function of $\tilde{\chi}(\boldsymbol r)$, we determine that $P_{\rm ave}(|\psi(\boldsymbol r)|^2)=P_{\rm TW}(\tilde{\chi}){\rm d}\tilde{\chi}/{\rm d}|\psi(\boldsymbol r)|^2$. Consequently, we obtain the following expression for the average wave density:
\begin{equation}
\langle|\psi(\boldsymbol r)|^2\rangle = e^{\langle|\ln\psi(\boldsymbol r)|^2\rangle}I(r),
\label{Gaussian_to_TW}
\end{equation}
with
\begin{equation}
\langle\ln|\psi(\boldsymbol r)|^2\rangle \displaystyle \underset{r\gg \xi}{\approx} -\frac{2r}{\xi}+\left(\frac{r}{\xi}\right)^{1/3} \Gamma \, \mu\; ,
\label{Eq:sm_fit_ave_log}
\end{equation}
and
\begin{equation}
I(r)=\int_{-\infty}^{\infty}e^{\sigma(r) x}P_{\rm TW}(x){\rm d}x.
\label{Eq:integrand}
\end{equation}
Hence, the average wave density can be expressed as the typical wave density multiplied by an additional $r$-dependent integral $I(r)$, {where $\langle...\rangle$ denotes disorder averaging at fixed $r$, $\mu\approx -1.77$ is the non-zero mean of the GUE TW distribution and $\sigma(r)$ represents $\sigma[\langle|\ln\psi(r)|^2\rangle]$.} The integral $I(r)$ is a bilateral Laplace transform of the 
rescaled Tracy-Widom distribution, and since $\sigma(r)$ is a large positive number, it
is primarily influenced by the extreme values in the right tail of the rescaled TW distribution, i.e. $P_{\rm TW}(x)$ at large $x$. The right tail of the GUE TW distribution is usually approximated as $P_{\rm TW}(x) \sim \exp(-\alpha x^{3/2})$ with $\alpha$ a constant, however here, to facilitate analytical tractability, we will use Laplace's method and approximate the right tail of the TW distribution function (for $x>0$) using the following expression:
\begin{equation}
P_{\rm TW}(x)\approx e^{-Ax^2+Bx+C}.
\label{Eq:tw_approx}
\end{equation}
By fitting the rescaled GUE TW distribution, we determine the parameters $A=0.18$, $B=-1.42$, and $C=1.10$ within the range $x\in [2,12]$, {demonstrated in Fig.~\ref{fig:average_wave_den}(a).} Consequently, we obtain the expression for the average wave density as follows:
\begin{eqnarray}
I_{\rm G}(r) 
&=& \int_{0}^{\infty}e^{-Ax^2+(\sigma(r)+B)x+C}{\rm d}x \nonumber\\
&=& \frac{\sqrt{\pi}e^{C+\frac{(\sigma(r)+B)^2}{4A}}(1+{\rm erf}(\frac{\sigma(r)+B}{2\sqrt{A}}))}{2\sqrt{A}} \nonumber\\
&\approx& \sqrt{\frac{\pi}{A}}e^{C+\frac{(\sigma(r)+B)^2}{4A}},
\label{intergral_G}
\end{eqnarray}
where ${\rm erf}$ is the error function, and we have made use of the approximation ${\rm erf}(\frac{\sigma(r)+B}{2\sqrt{A}})\approx 1$ for sufficiently large positive $\sigma(r)$. Applying these, we obtain the following expression:
\begin{equation}
I(r) \approx I_{\rm G}(r) \approx \sqrt{\frac{\pi}{A}} e^{C+\frac{B^2}{4A}} e^{\frac{BD}{2A}(r/\xi)^{1/3}+\frac{D^2}{4A}(r/\xi)^{2/3}},
\label{ave_den_integration}
\end{equation}
{where we have replaced $\sigma(r) = D(r/\xi)^{1/3}$ with $D\approx 1.39$, presented in Fig.~\ref{fig:average_wave_den}(b).}

{In Fig.~\ref{fig:average_wave_den}(c), we present our numerical results justifying our approximation to the integral $I(r)$ by $I_{\rm G}(r)$ in Eq.~\eqref{ave_den_integration}.}
An excellent agreement is observed for a wide range of $r$. Finally, by plugging {the analytical form of the typical wave density Eq.~\eqref{Eq:sm_fit_ave_log}} and Eq.~\eqref{ave_den_integration} into Eq.~\eqref{Gaussian_to_TW}, we obtain
\begin{eqnarray}
\langle|\psi(\boldsymbol r)|^2\rangle &\approx& \sqrt{\frac{\pi}{A}} e^{C+\frac{B^2}{4A}} e^{-2r/\xi+(\Gamma\mu+\frac{BD}{2A})(r/\xi)^{1/3}+\frac{D^2}{4A}(r/\xi)^{2/3}}\nonumber\\
&=& e^{-{2r}/{\xi}+\Gamma^\prime({r}/{\xi})^{1/3}+\Gamma^{\prime\prime}({r}/{\xi})^{2/3}+\Gamma^{\prime\prime\prime}},
\label{Eq:sm_approx_ave_den}
\end{eqnarray}
with $\Gamma^\prime=\Gamma\mu+\frac{BD}{2A}=-6.8,\Gamma^{\prime\prime}=\frac{D^2}{4A}=2.7$ and $\Gamma^{\prime\prime\prime}=\ln{\sqrt{\frac{\pi}{A}}}+\frac{B^2}{4A}+C=5.3$. This expression offers an accurate approximation of the average wave density.

\section{2D Anderson model}

In this section, we present additional numerical results concerning localized wave packets generated using the Anderson model on a 2D square lattice. The Hamiltonian of the Anderson model is written as:
\begin{equation}
H_{\rm AL} = -J\sum_{\langle \boldsymbol r,\boldsymbol r^\prime\rangle}(c^\dagger_{\boldsymbol r}c_{\boldsymbol r^\prime}+c^\dagger_{\boldsymbol r^\prime} c_{\boldsymbol r})+\sum_{\boldsymbol r} w_{\boldsymbol r}n_{\boldsymbol r},
\label{H_AL}
\end{equation}
Here, $c^\dagger_{\boldsymbol r}$ and $c_{\boldsymbol r}$ represent the creation and annihilation operators, respectively, and $n_{\boldsymbol r}=c^\dagger_{\boldsymbol r}c_{\boldsymbol r}$ is the number operator on the lattice site $\boldsymbol r=(x,y)$. We work under periodic boundary conditions. The notation $\langle \boldsymbol r,\boldsymbol r^\prime\rangle$ denotes nearest neighbors, $J=1$ is the hopping amplitude, and $w_{\boldsymbol r} \in [-W,W]$ represents the onsite disorder uniformly drawn from a box distribution with strength $W>0$. The temporal evolution of a wave packet initially taken as $|\psi(t=0)\rangle=|\psi_0\rangle$ given by Eqs.~(2) or (3) of the main text, is given by:
\begin{equation}
|\psi_t\rangle=e^{-iH_{\rm AL}t}|\psi_0\rangle
\end{equation}
where we have set $\hbar=1$.

\begin{figure}
\includegraphics[width=0.95\linewidth]{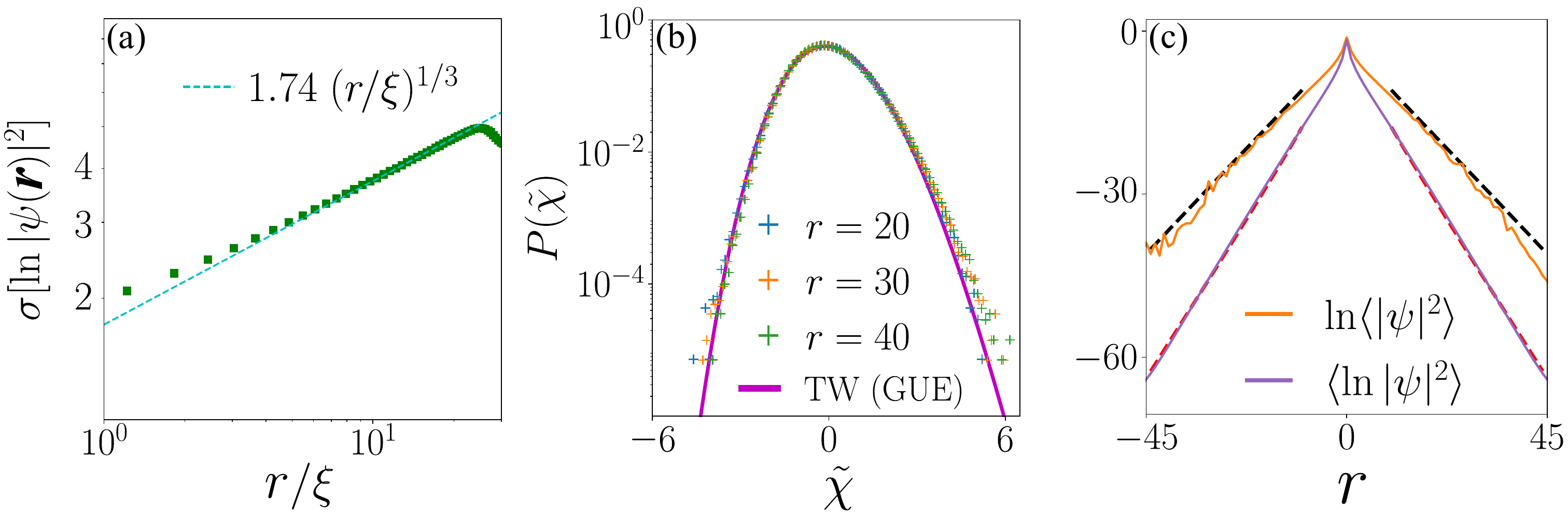}
\caption{{Localized wave packets in the Anderson model. (a) Scaling of the fluctuation of $\ln|\psi(\boldsymbol r)|^2$ with the distance $r$. The cyan dashed line indicates the KPZ expected behavior $\sigma[\ln|\psi(\boldsymbol r)|^2] \approx D(r/\xi)^{1/3}$, with $D=1.74$ and $\xi=1.6$ the localization length. (b) Normalized distribution of the rescaled variable $\tilde{\chi}(\boldsymbol r)=(\ln|\psi(\boldsymbol r)|^2-\langle\ln|\psi(\boldsymbol r)|^2\rangle)/{\sigma[\ln|\psi(\boldsymbol r)|^2]}$ with different $r$ represented by different colors, alongside the rescaled GUE TW distribution (solid magenta line). (c) Numerical results (purple solid line) for the typical wave density $\langle\ln|\psi(\boldsymbol r)|^2\rangle$ and Eq.~\eqref{Eq:sm_fit_ave_log} (red dashed line) with fitting parameters $\xi=1.6$ and $\Gamma=1.7$. Numerical results (orange solid line) for the average wave density $\ln\langle|\psi(\boldsymbol r)|^2\rangle$ and Eq.~\eqref{Eq:sm_approx_ave_den} (black dashed line) with fitting parameters $\Gamma^\prime\approx -9.9$, $\Gamma^{\prime\prime}\approx -4.2$ and $\Gamma^{\prime\prime\prime}\approx 5.3$. In the numerical simulations, we have considered system of size $101\times 101$, hopping amplitude $J=1$, disorder strength $W=8$, evolution time $t=500$ and $10^6$ disorder realizations.} 
}
\label{fig:anderson_std_tw_wave}
\end{figure}

We have investigated the scaling behaviours of the fluctuations of $\ln|\psi(\boldsymbol r)|^2$ with respect to the distance $r$. We recover that the fluctuation exponent $\beta \approx 1/3$ for the standard deviation $\sigma[\ln|\psi(\boldsymbol r)|^2] \sim r^{\beta}$ as illustrated in Fig.~\ref{fig:anderson_std_tw_wave}(a). 
The distribution function of the variable $\tilde{\chi}(\boldsymbol r)=(\ln|\psi(\boldsymbol r)|^2-\langle\ln|\psi(\boldsymbol r)|^2\rangle)/\sigma[\ln|\psi(\boldsymbol r)|^2]$ is well described by the rescaled Tracy-Widom distribution, as depicted in Fig.~\ref{fig:anderson_std_tw_wave}(b). 
Last, the spatial profiles of the typical and average wave densities in Fig.~\ref{fig:anderson_std_tw_wave}(c), are well fitted by their respective analytical forms, Eq.~(9) and Eq.~(10) of the main text. These results further confirm that the critical exponents and statistical distributions governing KPZ physics are universally present in localized wave packets in two dimensions.

\bibliography{reference_glassy_qkr.bib}

\end{document}